\documentclass[a4paper, amsmath, amssymb, 11pt]{article}

\usepackage[top=2cm, bottom=2cm, left=3cm, right=2cm]{geometry}
\usepackage[fleqn]{amsmath}
\usepackage{amstext,amsbsy,amscd,amssymb,amsfonts}
\begin{document}

\begin{center}
\large {\bf Asymptotics of Few-body Equations of Collision Theory}\\
\vspace{0.16cm}
{\large Vagner Jikia, Jemal Mebonia \\
{\small I. Javakhishvili Tbilisi State University, Tbilisi, Georgia,\\ Email: v\_ jikia@yahoo.com}} 

\end{center}

\vspace{0.2cm}
{\bf Abstract}
A correct high-energy asymptotic form of Faddeev type few-body integral equations is found. Iterative series corresponding to these asymptotic relations converge with a certain accuracy to a finite sum, which satisfies the corresponding condition of unitarity.

\vspace{0.4cm}
{\bf Introduction.}
It is well known that in the case of three or more particles the Lipmann-Schwinger (LS) integral equation is not of Fredholm type. This is the consequence of the existence of bounded subsystems. By using identical transformations of the kernel of LS equation for certain classes of potentials Faddeev obtained Fredholm type integral equations for the three-body problem $[1]$. The motion of for bodies is described by equations of Yakubovsky and Alt-Grassberger-Sandhas-Khelashvili $[2.3]$, which are obtained as a result of two subsequent transpormations of the kernel of LS equation. For the case $N>4$, it has been shown that the kernel of the corresponding integrals equations after $N - 2$ iterations became compact on the plane of complex energy variable $z$, if ${\mathop{\rm Im}\nolimits} z \ne 0$.   For ${\mathop{\rm Im}\nolimits} z = 0$ the compactness has not been proven yet $[4]$. In turned out that, in the case of $N \ge 3,$ the mentioned equations have correct asymptotic form the derivation and detailed analysis of which is presented this paper.\\ 


\begin{center}
\large {\bf Few-body equations, general formalism}
\end{center}

Let us consider a system of $N$ particles in non-relativistic approximation. Interaction operator $V$ in the approximation of two-particle forces is given as:   
\begin{eqnarray}
V = \sum\limits_\alpha  {{\upsilon _\alpha }} ,\quad \alpha  \equiv \left( {mn} \right);\quad m,n = 1,2,\;.\;.\;.\;,N;\;\,m < n,N \ge 3.
\end{eqnarray}
Here ${\upsilon _\alpha }$ are potentials of pair-interactions.

$N$-particle scattering $T$-matrix is determined by the LS equation:
\begin{eqnarray} 
T\left( z \right) = V + V{G_0}\left( z \right)T\left( z \right).
\end{eqnarray}
Here ${G_0}$ is the Green's function of $N$ free particles:
\begin{eqnarray} 
{G_0}\left( z \right) = {\left( {z - {H_0}} \right)^{ - 1}},\quad z = {E_0} + i\varepsilon ,
\end{eqnarray}
where ${H_0}$ is the free Hamiltonian of the system and ${E_0}$ is the corresponding eigenvalue. ${G_0}$ can be presented as a sum of anti-Hermitian and Hermitian parts [5]:  
\begin{eqnarray} 
&{G_0}\left( z \right) = {G_1}\left( z \right) + {G_2}\left( z \right),\\
&{G_1}\left( z \right) = \frac{1}{2}\left( {{G_0}\left( z \right) - {G_0}\left( {\tilde z} \right)} \right),\quad {G_2}\left( z \right) = \frac{1}{2}\left( {{G_0}\left( z \right) + {G_0}\left( {\tilde z} \right)} \right).
\end{eqnarray}
Here $\tilde z$ stands for the complex conjugate of the $z$ parameter. 

To solve equation (2) formally one applies the standard Faddeev approach and obtains: 
\begin{eqnarray}
T\left( z \right) = \sum\limits_\alpha  {{T^\alpha }\left( z \right)}, 
\end{eqnarray}
where auxillary operators ${T^\alpha }$ are define by the following system of equations: 
\begin{eqnarray} 
{T^\alpha }\left( z \right) = {T_\alpha }\left( z \right) + {T_\alpha }\left( z \right){G_0}\left( z \right)\sum\limits_{\beta  \ne \alpha } {{T^\beta }} \left( z \right),
\end{eqnarray}
and for two-particle operators ${T_\alpha }$ one can write:
\begin{eqnarray}
{T_\alpha }\left( z \right) = {\upsilon _\alpha } + {\upsilon _\alpha }{G_0}\left( z \right){T_\alpha }\left( z \right).
\end{eqnarray}
Equation (2) can be also solved formally by applying the Heitler formalism of separation [6] resulting in:
\begin{eqnarray}
&T\left( z \right) = K\left( z \right) + K\left( z \right){G_1}\left( z \right)T\left( z \right),\\
&K\left( z \right) = V + V{G_2}\left( z \right)K\left( z \right).
\end{eqnarray}
From the Heitler equation (9) $T$ can be expressed in terms of a Hermitian operator $K$ which in turn is determined by Eq. (10). If Eqs. (9) and (10) are correct equations (the two-particle case) then the $T$ operator constructed this way satisfies the corresponding condition of unitarity [6] for any approximation of $K.$ 

Similarly to Eq. (2) let us look for the solution of Eq. (10) in the form of a sum:
\begin{eqnarray}
K\left( z \right) = \sum\limits_\alpha  {{K^\alpha }\left( z \right)}, 
\end{eqnarray}
where auxiliary operators ${K^\alpha }$ are given by equation:
\begin{eqnarray}
{K^\alpha }\left( z \right) = {K_\alpha }\left( z \right) + {K_\alpha }\left( z \right){G_2}\left( z \right)\sum\limits_{\beta  \ne \alpha } {{K^\beta }} \left( z \right),
\end{eqnarray}
and for two particle operators $ {K_\alpha }$ we have: 
\begin{eqnarray}
{K_\alpha }\left( z \right) = {\upsilon _\alpha } + {\upsilon _\alpha }{G_2}\left( z \right){K_\alpha }\left( z \right).
\end{eqnarray}
It is well known that Eqs. (7) and (12) are not compact in general.
 
Below we show that by using the Heitler formalism it is possible to find the correct asymptotic (high energy) form of the system of multi-particle integral equations (7).  

\begin{center}
\large {\bf Multi-particle equations in asymptotics}
\end{center}

Using the behaviour of Eq. (12) at sufficiently high energies:  ${\mathop{\rm Re}\nolimits} \left( z \right) \gg E_{\min }^B $ ($ E_{\min }^B $ denotes the absolute magnitude of the minimum of all allowed binding energies in the system to be considered), let us restrict Eq. (9) to its high-energy approximation and using the obtained result let us rewrite Eq. (7) in asymptotics.  

It is well-known that the square of the kernel of Eq. (12) can be expressed in terms of the following form: 
\begin{eqnarray}
{K_\alpha }\left( z \right){G_2}\left( z \right){K_\beta }\left( z \right){G_2}\left( z \right),\quad \alpha  \ne \beta ,
\end{eqnarray}
which in the approximation of Eq. (38) (see appendix) can be negleqted. This is why the kernel of Eq. (12) at sufficiently high energies is square integrable so that the corresponding Neuman series converges. To obtain the desired system of equations let us keep the first order asymptotic terms in Eq. (12) ($ K $-matrix impulse approximation, see [7]): 
\begin{eqnarray}
{K^\alpha }\left( z \right) = {K_\alpha }\left( z \right),\quad K\left( z \right) = \sum\limits_\alpha  {{K_\alpha }\left( z \right)}.
\end{eqnarray}
We substitute the $ K $ operator of Eqs. (15) into Eq. (9) and obtain:
\begin{eqnarray}
T\left( z \right) = \sum\limits_\alpha  {{K_\alpha }\left( z \right)}  + \sum\limits_\alpha  {{K_\alpha }\left( z \right)} {G_1}\left( z \right)T\left( z \right).
\end{eqnarray}
Equation (16) represents the Heitler equation in the $ K $-matrix impulse approximation. It is defined in the kinematic area: $
{\mathop{\rm Re}\nolimits} \left( z \right) \gg E_{\min }^B. $
In the approximation of Eq. (1) the Lippmann-Shwinger equation and Eq. (16) have the same form and therefore one can use the standard well-known approach by Faddeev for the latter.  

Let us show that, using the Faddeev's method, Eq. (16) reduces to the desired asymptotic form. For this purpose we decompose $ T $ in the form of Eq. (6) where, according to Eq. (16) 
\begin{eqnarray}
{T^\alpha }\left( z \right) = {K_\alpha }\left( z \right) + {K_\alpha }\left( z \right){G_1}\left( z \right)T\left( z \right).
\end{eqnarray}
The insertion of Eq. (6) into (17) gives: 
\begin{eqnarray}
{T^\alpha }\left( z \right) = {K_\alpha }\left( z \right) + {K_\alpha }\left( z \right){G_1}\left( z \right)\sum\limits_{\beta  = \alpha } {{T^\beta }\left( z \right)} .
\end{eqnarray}
To exclude the diagonal terms from Eq. (18) let as rewrite it in the following form:
\begin{eqnarray}
\left( {1 - {K_\alpha }\left( z \right){G_1}\left( z \right)} \right){T^\alpha }\left( z \right) = {K_\alpha }\left( z \right) + {K_\alpha }\left( z \right){G_1}\left( z \right)\sum\limits_{\beta  \ne \alpha } {{T^\beta }\left( z \right)} .
\end{eqnarray}
Inverting $ 1 - {K_\alpha }\left( z \right){G_1}\left( z \right) $ operators in the expression (19) we obtain:
\begin{eqnarray}
{T^\alpha }\left( z \right) = {\left( {1 - {K_\alpha }\left( z \right){G_1}\left( z \right)} \right)^{ - 1}}{K_\alpha }\left( z \right) + {\left( {1 - {K_\alpha }\left( z \right){G_1}\left( z \right)} \right)^{ - 1}}{K_\alpha }\left( z \right){G_1}\left( z \right)\sum\limits_{\beta  \ne \alpha } {{T^\beta }\left( z \right)} .
\end{eqnarray}
Using two-particle equations:
\begin{eqnarray}
{T_\alpha }\left( z \right) = {K_\alpha }\left( z \right) + {K_\alpha }\left( z \right){G_1}\left( z \right){T_\alpha }\left( z \right),
\end{eqnarray}
one can write:
\begin{eqnarray}
{K_\alpha }\left( z \right) = {\left( {1 + {T_\alpha }\left( z \right){G_1}\left( z \right)} \right)^{ - 1}}{T_\alpha }\left( z \right).
\end{eqnarray}
Inserting Eq. (22) into Eq. (20) we obtain:
\begin{eqnarray}
{T^\alpha }\left( z \right) = {T_\alpha }\left( z \right) + {T_\alpha }\left( z \right){G_1}\left( z \right)\sum\limits_{\beta  \ne \alpha } {{T^\beta }\left( z \right)}.
\end{eqnarray}
Eq. (23) represents asymptotic $
\left( {{\mathop{\rm Re}\nolimits} z \gg E_{\min }^B} \right) $
form of the system (7) of $ N $-particle integral equations, which indicates that at high energies in the Green's function $ {G_0} $ entering in Eq. (7) its Hermitian part can be negleqted (for the two-particle case see Ref. [8]). Equation (23) is a system of equations for auxiliary operators $ {T^\alpha } $ in the $ K $-matrix impulse approximation. Using the estimation of Eq. (40) (see appendix) one can  write: 
\begin{eqnarray}
\left\| {{T_\alpha }\left( z \right){G_1}\left( z \right){T_\beta }\left( z \right){G_1}\left( z \right)} \right\| = O\left( {{\varepsilon ^2}} \right) \approx 0,\quad \alpha  \ne \beta ,
\end{eqnarray}
Iterative series corresponding to Eq. (23) converges with the accuracy, specified in Eq. (24), to the following finite sum:
\begin{eqnarray}
{T^\alpha }\left( z \right) = {T_\alpha }\left( z \right) + {T_\alpha }\left( z \right){G_1}\left( z \right)\sum\limits_{\beta  \ne \alpha } {{T_\beta }\left( z \right)} ,\quad {\mathop{\rm Re}\nolimits} \left( z \right) \gg E_{\min }^B.
\end{eqnarray}
The specified accuracy of convergence guarantees the unitarity of $ T $-matrix [9].

Equation (23) describes scattering on weekly bound system
$ \left( {{\mathop{\rm Re}\nolimits} \left( z \right) \gg E_{\min }^B} \right). $ According to the approximation of Eq. (15), equation (23) corresponds to the processes in which single collision dominate. The expression of the kernel of the obtained asymptotic equations with the Green's function $ {G_1} $ indicates that these equations are adequate for elastic and quasi-elastic scattering reactions (the internal energy of the system does not change). 

Let us insert Eq. (25) in Eq. (6):
\begin{eqnarray}
&T\left( z \right) = \sum\limits_\alpha  {{T_\alpha }\left( z \right)\left( {1 + {G_1}\left( z \right)\sum\limits_{\beta  \ne \alpha } {{T_\beta }\left( z \right)} } \right),\quad } {\mathop{\rm Re}\nolimits} \left( z \right) \gg E_{\min }^B,\\
&\alpha ,\beta  \equiv \left( {mn} \right),\quad \;m,n = 1,2,\;.\;.\;.\;,N;\;\,m < n,N \ge 3\nonumber.		  
\end{eqnarray}
Equation (26) is a particular unitary solution of Eq. (16), which at the same time represents an asymptotic solution of Eqs. (2) and (9).

\begin{center}
\large {\bf Summary}\\
\end{center}

Thus, by using the Heitler formalism in multi-particle problem we got multi-particle asymptotic equations. In particular, keeping in the right-hand side of Eq. (9) the high-energy terms we obtained asymptotic equation (16). Based on this equation we constructed the Faddeev type compact the system of $ N $-particle integral equations in asymptotic, Eq. (23). Using the convergence of the iterative series corresponding to Eq. (23) we obtained Eq. (26), the unitary asymptotic solution of multi-particle Eqs. (2) and (9), which is identical to the result proposed by us earlier [9].   
\\
\\
{\bf Acknowledgments} The authors thank J. Gegelia, A. Khelashvili, A. Kvinikhidze, I. Lomidze and M. Makhviladze for discussions and comments on the article. 

\begin{center}
\large {\bf Appendix}
\end{center}

Below we establish the high-energy behavior of operators determined by Eqs. (8) and (13). The obtained result is used for the estimation of the kernel of multi-particle equations.

Let as show that
\begin{eqnarray}
{T_\alpha }\left( z \right) \approx {\upsilon _\alpha },\quad {\mathop{\rm Re}\nolimits} \left( z \right) > {\mathop{\rm Re}\nolimits} \left( {\bar z} \right).
\end{eqnarray} 
Here $ z = {E_0} + i\varepsilon . $ $ {E_0} $ is a kinetic energy of the considered system. $ \bar z = \bar E + i\varepsilon . $ $ \bar E $
is the fixed energy, which satisfies the condition: $ \bar E \gg E_{\min }^B. $ $ E_{\min }^B $ denotes the absolute value of the minimum of all allowed binding energies in the considered system.

Using the symbolic notation $ \alpha  \equiv \left( {mn} \right), $ we write the connection among $ {t_\alpha } $ and $ {T_\alpha } $ operators:
\begin{eqnarray}
\left\langle {{{q'}_1}} \right.\,{q'_2}\,.\;\,.\,\;{q'_N}\left| {{T_\alpha }\left( z \right)} \right|\left. {{q_1}{q_2}\,.\;\,.\,\;{q_N}} \right\rangle  = \prod\limits_{l \ne m \ne n \ne l}^N {\delta \left( {{{q'}_l} - {q_l}} \right)} \left\langle {{{q'}_m}} \right.\,{q'_n}\,\left| {{t_\alpha }\left( {z'} \right)} \right|\left. {{q_m}{q_n}} \right\rangle ,
\end{eqnarray}
where the pure two particle interaction operator $ {t_\alpha } $ satisfies the LS integral equation:
\begin{eqnarray}
{t_\alpha }\left( {z'} \right) = {\upsilon _\alpha } + {\upsilon _\alpha }{g_0}\left( {z'} \right){t_\alpha }\left( {z'} \right).
\end{eqnarray}
Here $ {g_0} $ is the Greenq's function of two free particles.
\begin{eqnarray}
z' = E' + i\varepsilon ,\quad E' = {E_0} - \sum\limits_{n \ne l \ne m} {\frac{{q_l^2}}{{2{m_l}}}} .
\end{eqnarray}
$ {E_0} $ and $ {q_l} $ are fixed parameters. According to the Klein-	Zemach theorem [10] we have:
\begin{eqnarray}
{t_\alpha }\left( {z'} \right) \approx {\upsilon _\alpha },\quad {\mathop{\rm Re}\nolimits} \left( {z'} \right) > {\mathop{\rm Re}\nolimits} \left( {\bar z} \right).
\end{eqnarray}
Inserting the relation (31) into Eq. (28) we get Eq. (27).

Let us express $ {T_\alpha } $ from Eq. (8) and substitute the obtained result in Eq. (27):
\begin{eqnarray}
{T_\alpha }\left( z \right) = {\left( {1 - {\upsilon _\alpha }{G_0}\left( z \right)} \right)^{ - 1}}{\upsilon _\alpha } \approx {\upsilon _\alpha },\quad {\mathop{\rm Re}\nolimits} \left( z \right) > {\mathop{\rm Re}\nolimits} \left( {\bar z} \right).
\end{eqnarray}
Due to Eqs. (27) and (32) we obtained:
\begin{eqnarray}
\left\| {{T_\alpha }\left( z \right){G_0}\left( z \right)} \right\| \approx \left\| {{\upsilon _\alpha }{G_0}\left( z \right)} \right\| \approx \varepsilon ,\quad {\mathop{\rm Re}\nolimits} \left( z \right) > {\mathop{\rm Re}\nolimits} \left( {\bar z} \right).
\end{eqnarray}

Similarly to Eq. (28) we write:
\begin{eqnarray}
\left\langle {{{q'}_1}} \right.\,{q'_2}\,.\;\,.\,\;{q'_N}\left| {{K_\alpha }\left( z \right)} \right|\left. {{q_1}{q_2}\,.\;\,.\,\;{q_N}} \right\rangle  = \prod\limits_{l \ne m \ne n \ne l}^N {\delta \left( {{{q'}_l} - {q_l}} \right)} \left\langle {{{q'}_m}} \right.\,{q'_n}\,\left| {{k_\alpha }\left( {z'} \right)} \right|\left. {{q_m}{q_n}} \right\rangle .
\end{eqnarray}
Here $ {k_\alpha } $ is a two particle Hermitian operator:
\begin{eqnarray}
{k_\alpha }\left( {z'} \right) = {\upsilon _\alpha } + {\upsilon _\alpha }{g_2}\left( {z'} \right){k_\alpha }\left( {z'} \right),
\end{eqnarray}
where $ {g_2} $ can be expressed as follows:
\begin{eqnarray}
{g_2}\left( {z'} \right) = \frac{1}{2}\left( {{g_0}\left( {z'} \right) + {g_0}\left( {\tilde z'} \right)} \right).
\end{eqnarray}
The parameter $ z' $ is defined from equalities (30).

Due to the mentioned Klein-Zemach theorem, one can show similarly to Eq. (27):
\begin{eqnarray}
{K_\alpha }\left( z \right) \approx {\upsilon _\alpha },\quad {\mathop{\rm Re}\nolimits} \left( z \right) > {\mathop{\rm Re}\nolimits} \left( {\bar z} \right).
\end{eqnarray}
According to Eqs. (13) and (37) we have:
\begin{eqnarray}
\left\| {{K_\alpha }\left( z \right){G_2}\left( z \right)} \right\| \approx \left\| {{\upsilon _\alpha }{G_2}\left( z \right)} \right\| \approx \varepsilon ,\quad {\mathop{\rm Re}\nolimits} \left( z \right) > {\mathop{\rm Re}\nolimits} \left( {\bar z} \right).
\end{eqnarray}
From Eq. (21) in the approximation of Eq. (37) we obtain [8]:
\begin{eqnarray}
{T_\alpha }\left( z \right) = {\upsilon _\alpha } + {\upsilon _\alpha }{G_1}\left( z \right){T_\alpha }\left( z \right).
\end{eqnarray}
Using Eqs. (39) and (27) one can write:
\begin{eqnarray}
\left\| {{T_\alpha }\left( z \right){G_1}\left( z \right)} \right\| \approx \left\| {{\upsilon _\alpha }{G_1}\left( z \right)} \right\| \approx \varepsilon ,\quad {\mathop{\rm Re}\nolimits} \left( z \right) > {\mathop{\rm Re}\nolimits} \left( {\bar z} \right).
\end{eqnarray}

Thus, it is shown that in the asymptotic $ \left( {{\mathop{\rm Re}\nolimits} \left( z \right) \gg E_{\min }^B} \right), $ the kernels of multi-particle equations are sufficiently small, so that the corresponding iterative series converge.

\end{document}